\documentclass[conference, twoside]{IEEEtran}
\IEEEoverridecommandlockouts

\usepackage{cite}
\usepackage{amsmath,amssymb,amsfonts, amsthm}
\usepackage{algorithm}
\usepackage{algpseudocode}
\usepackage{graphicx}
\graphicspath{{Fig/}{../Fig/}{fig/}{../fig}}
\usepackage{textcomp}
\usepackage{xcolor}
\usepackage{color}
\usepackage{enumitem}
\usepackage{adjustbox}

\providecommand{\com[2]}{\begin{tt}[#1: #2]\end{tt}}
\definecolor{red}{rgb}{1,0,0}

\definecolor{blu}{rgb}{0,0,1}

\definecolor{gre}{rgb}{0,0.7,0.3}

\newcommand{\bydef}{\stackrel{\Delta}{=}}

\newcommand{\beq}{\begin{equation}}
\newcommand{\eeq}{\end{equation}}
\newcommand{\beqa}{\begin{eqnarray}}
\newcommand{\eeqa}{\end{eqnarray}}
\newcommand{\beqan}{\begin{eqnarray*}}
\newcommand{\eeqan}{\end{eqnarray*}}
\newcommand{\bef}{\begin{figure}}
\newcommand{\enf}{\end{figure}}

\newcommand{\del}[1]{}

\definecolor{magenta}{cmyk}{0.5, 1, 0, 0}
\definecolor{greenedu}{cmyk}{1, 0, 1, 0.1}
\definecolor{cyanedu}{cmyk}{1, 0, 0, 0.1}

\newcommand{\bi}{\begin{itemize}}
\newcommand{\ei}{\end{itemize}}
\newcommand{\bc}{\begin{center}}
\newcommand{\ec}{\end{center}}
\newcommand{\ba}{\begin{array}}
\newcommand{\ea}{\end{array}}
\newcommand{\be}{\begin{equation}}
\newcommand{\ee}{\end{equation}}
\newcommand{\beno}{\begin{equation*}}
\newcommand{\eeno}{\end{equation*}}
\newcommand{\beqna}{\begin{eqnarray}}
\newcommand{\eeqna}{\end{eqnarray}}
\newcommand{\bd}{\begin{displaymath}}
\newcommand{\ed}{\end{displaymath}}
\newcommand{\beqnd}{\begin{eqnarray*}}
\newcommand{\eeqnd}{\end{eqnarray*}}

\renewcommand{\ni}{\noindent}

\newcommand{\cqfd}{\hfill \rule{2mm}{2mm}\smallbreak\indent}

\newtheorem{definition}{Definition}
\newtheorem{theorem}{Theorem}
\newtheorem{lemma}{Lemma}

\newtheorem{proposition}{Proposition}

\numberwithin{definition}{section}
\numberwithin{theorem}{section}
\numberwithin{lemma}{section}
\numberwithin{corollary}{section}
\numberwithin{example}{section}
\numberwithin{proposition}{section}

\newcommand{\emp}[1]{\ensuremath{\text{EMP}_{#1}}}

\title{Excitation and Measurement Patterns for the Identifiability of Directed Acyclic Graphs }
\author{Eduardo Mapurunga, Michel Gevers, \IEEEmembership{Life Fellow, IEEE,} and Alexandre S. Bazanella, \IEEEmembership{Senior Member, IEEE} 
\thanks{Eduardo Mapurunga and Alexandre S.  Bazanella are with the Data Driven Control Group, Department of Automation and Energy, 
Universidade Federal do Rio Grande do Sul, Porto Alegre-RS, Brazil, \{eduardo.mapurunga, bazanella\}@ufrgs.br}
\thanks{Michel Gevers is with the Institute of Information and Communication Technologies, Electronics and Applied Mathematics (ICTEAM), UCLouvain, Louvain la Neuve, Belgium,
michel.gevers@uclouvain.be.}
\thanks{This work  was supported in part by the Coordena{\c c}{\~ a}o de Aperfei{\c c}oamento de Pessoal de N{\'i}vel Superior - Brasil (CAPES) - Finance Code 001, by Conselho Nacional de Desenvolvimento Cient{\' i}fico e Tecnol{\' o}gico (CNPq),
by Wallonie-Bruxelles International (WBI), by a WBI.World excellence fellowship, and by a Concerted Research Action (ARC) of the French Community of Belgium.}
}

\begin{document}

\maketitle

%
%
\begin{abstract}                
 This paper deals with the design of Excitation and Measurement Patterns (EMP) for the identification of a class of dynamical networks whose topology has the structure of a Directed Acyclic Graph (DAG). In addition to the by now well known condition that the identifiabiltiy of any dynamical network requires that the sources be excited, the sinks be measured, and all other nodes be either excited or measured, we show that for DAGs two other types of nodes have special excitation and measurement requirements. Armed with this result, we propose a systematic  procedure for the design of EMPs that guarantee identifiability of a network with DAG topology.
\end{abstract}


\section{Introduction}

This work deals with identifiability of dynamic networks, which has been an active research topic in the control community over the last ten years. The network framework used here was introduced in \cite{vandenhof-dankers-heuberger-etal-identification-2013} where signals were represented as nodes of the network which were related to other nodes through transfer functions.  These networks can be interpreted as directed graphs where the transfer functions, also called modules, are the edges of the graph and the node signals are the vertices.

In \cite{vandenhof-dankers-heuberger-etal-identification-2013} it was assumed that all nodes are excited and measured. As a result, an input-output matrix of the network,  denoted $T(z)$, can be defined, which can always be identified from these data. The network identifiability question is then whether the network matrix, denoted $G(z)$  (whose elements are the transfer functions relating the nodes) can be recovered from this closed-loop transfer matrix $T(z)$. 
In subsequent work, a range of new objectives were defined, from the  identification of the whole network  to identification of some specific part of the network 
\cite{vandenhof-dankers-heuberger-etal-identification-2013, gevers-bazanella-parraga-identifiability-2017, bazanella-gevers-hendrickx-etal-identifiability-2017, shi-cheng-vandenhof-generic-2022 , hendrickx-gevers-bazanella-identifiability-2019, everitt-bottegal-hjalmarsson-empirical-2018, gevers-bazanella-dasilva-practical-2018, van_waarde_necessary_2019, jahandari-materassi-sufficient-2021}. 
As for the assumptions on the signals, up to 2019, all contributions  assumed that either all nodes are excited, or all nodes are measured.  A typical question would be: given that all nodes are excited, which nodes must be measured in order to identify the whole network? 

The first identifiability results for networks where not all nodes are excited AND not all nodes are measured were presented in \cite{bazanella-gevers-hendrickx-network-2019}. That paper first provided a necessary condition for identifiability of any network: each node must be either excited or measured, at least one node must be excited and at least one node measured.  The paper \cite{bazanella-gevers-hendrickx-network-2019} also  presented identifiability conditions for two special classes of networks, namely trees and loops. 

The results of  \cite{bazanella-gevers-hendrickx-network-2019} inspired the definition of an excitation and measurement pattern (EMP), namely the combination of excited nodes and measured nodes.    The concept of EMP was introduced in \cite{mapurunga-optimal-2021} where an EMP was called {\it valid} if it guarantees  the identifiability of the whole network. An EMP  was called {\it minimal} if it  guarantees the identifiability of the network using the smallest possible combination of excited and measured nodes \cite{mapurunga-optimal-2021}. This number is the {\it cardinality of the EMP.} Achieving identifiability of a network with a minimal EMP is of both theoretical and practical interest. The excitation of a node typically requires an actuator, while its measurement requires a measurement device. On the other hand, having some flexibility in the choice of a valid EMP is also of practical interest. It may be that exciting node 42, say, is prohibitively expensive while its measurement is easy; conversely, measuring a node may be difficult while its excitation is cheap. In evaluating the choice of an EMP for the identification of a network, one must of course remember that each node must be either excited or measured, or both. As a result, the cardinality of a valid EMP is always at least equal to $n$, the number of nodes.

The search for valid, and possibly minimal, EMPs began by looking at special structures. In \cite{bazanella-gevers-hendrickx-network-2019} a necessary and sufficient condition was given for the identifiability of a tree, which shows that a tree can possibly be identified with an EMP of cardinality $n$.  
In \cite{mapurunga-identifiability-2021} necessary and sufficient conditions were derived for the identifiability of some classes of parallel networks. 
In \cite{mapurunga-gevers-bazanella-necessary-2022}  necessary and sufficient conditions were given for the identifiability of loops.  This result showed that any loop with more than 3 nodes can also be identified with a minimal EMP of cardinality $n$. In addition, it was shown that constructing  EMPs for loops is very easy and that the number of minimal EMPs grows very quickly with the number of nodes. 

In this paper, we generalize the results derived in  \cite{bazanella-gevers-hendrickx-network-2019} for the identification of trees to a much wider class of networks, namely those that have the structure of a Directed Acyclic Graph (DAG), i.e.  a directed  graph that has no cycles. DAGs have been widely studied in the literature \cite{bang-jensen-gutin-digraphs-2009}. A specific feature of a DAG is that the corresponding network matrix $G(z)$ can be rewritten in a lower triangular form via a relabeling of the nodes. The corresponding input-output transfer matrix $T(z)$ is then also lower-triangular and this greatly simplifies the relations between the elements $G_{ij}$ of the network matrix and the elements $T_{ij}$ of this transfer matrix.

The main contributions of this paper are as follows. First we provide an explicit solution for the elements $G_{ij}(z)$ as a function of elements of  the matrix $T(z)$ and elements of its inverse: $S(z)= T^{-1}(z)$, with the property that the elements $S_{ij}(z)$ of  $S(z)$ are expressed in terms of $T_{kl}$ only (i.e. they do not involve inverses of elements $T_{kl}$). Next we focus on the construction of valid EMPs for the identification of a DAG. We provide a necessary condition on the excitation and measurement of specific nodes within a DAG. Finally, we provide a simple procedure  for the construction of a  valid EMP for the identification of a DAG. 


The paper is organized as follows. In Section~\ref{sec:Problem} we introduce the notations, and recall the definitions of generic identifiability of a network, and of a valid, as well as a minimal, Excitation and Measurement Pattern. We also recall the main necessary condition for identifiability of any network. In Section~\ref{sec:DAG} we present networks that have the topology of Directed Acyclic Graphs and establish the key results that will allow one to identify such networks. 
In Section~\ref{neccondDAG} we establish  necessary conditions for the identifiability of a DAG. The results of Sections~\ref{sec:DAG} and ~\ref{neccondDAG}  allow us to propose, in Section \ref{EMP1DAG},  a recursive procedure for the construction of a valid EMP, i.e. one that secures identifiability of the network. Finally,  we present conclusions in Section~\ref{Conclusion}.

\section{Definitions,  Notations and Concepts}
\label{sec:Problem}

In this Section, we briefly state the identifiability problem for a dynamical network and recall the main necessary condition for the identifiability of any such network. We then define the concept of a valid Excitation and Measurement Pattern, namely a choice of excited nodes and measured nodes that guarantee identifiability. We also introduce the notations used throughout the paper.

We consider dynamic networks composed of $n$ nodes (or vertices) which represent internal scalar signals $\left\lbrace w_k(t) \right\rbrace$ for $k \in  \{1, 2, \dots n\}$.
These nodes are interconnected by discrete time transfer functions, represented by edges, which are entries of a \emph{network matrix} $G(z)$.
The dynamics of the network is given by the following equations:
\begin{subequations}
	\begin{align}
	w(t) &= G(z)w(t) + Br(t), \label{eq:dynet1} \\
	y(t) &= Cw(t), \label{eq:dynet2}
	\end{align}
\end{subequations}
where $w(t) \in \mathbb{R}^n$ is the node vector, $r(t) \in \mathbb{R}^m$ is the input and $y(t) \in \mathbb{R}^p$ is the network's output.
The matrix $B \in \mathbb{Z}_2^{n \times m}$, where $\mathbb{Z}_2 \triangleq \{0, 1\}$, 
is a binary selection matrix with a single $1$ and $n-1$ zeros  in each column; it selects the inputs affecting the nodes of the network.
Similarly, $C \in \mathbb{Z}_2^{p \times n}$ is a matrix with a single $1$ and $n-1$ zeros in each row that selects which nodes are measured.

We now introduce some definitions and notations concerning these dynamical networks and their network matrix $G(z)$. To each $G(z)$ we can associate a directed graph $\mathcal{G}$ defined by the tuple $(\mathcal{V}, \mathcal{E})$, where $\mathcal{V}$ is the set of vertices and $\mathcal{E} \subseteq \mathcal{V} \times \mathcal{V}$ is the set of edges. The graph $\mathcal{G}$ defines the topology of the network.
A particular transfer function $G_{ji}(z)$ of the network matrix is called an incoming edge of node $j$ and outgoing edge of node $i$.
For the graph $\mathcal{G}$ associated to the network matrix $G(z)$ we introduce the following notations.
\begin{itemize}
	\item $\mathcal{W}$ - the set of all $n$ nodes;
	\item $\mathcal{B}$ - the set of excited nodes, defined by $B$ in (\ref{eq:dynet1});
	\item $\mathcal{C}$ - the set of measured nodes, defined by $C$ in (\ref{eq:dynet2});
	\item $\mathcal{F}$ - the set of sources: nodes with no incoming edges;
	\item $\mathcal{S}$ - the set of sinks: nodes with no outgoing edges;	
	\item $\mathcal{I}$ - the set of internal nodes, i.e. nodes that are neither a source nor a sink: $\mathcal{I} \triangleq \mathcal{W}\backslash(\mathcal{F}\cup\mathcal{S})$;
	\item $\mathcal{N}_j^-$ -  the set of in-neighbors  of node $j$;
	\item $\mathcal{N}_j^+$ - the set of out-neighbors of node $j$.
\end{itemize}
Additionally, we introduce the following two types of nodes.
\begin{itemize}
\item A node $j$ is called a \textbf{dource} if  it has at least one out-neighbor  that is connected to all  in-neighbors of $j$;
   \item  A node $j$ is called a \textbf{dink} if it has at least one in-neighbor that connects to all  
  out-neighbors of $j$.
\end{itemize}

The following assumptions are made about the network matrix:
\begin{itemize}
	\item the diagonal elements are zero;
	\item $(I - G(z))^{-1}$ is proper and stable.
\end{itemize}
One can represent the dynamic network in (\ref{eq:dynet1})-(\ref{eq:dynet2}) as an input-output model as follows
\begin{align}
	y(t) = M(z) r(t),\; \text{with}\; M(z) \triangleq C T(z) B.
	\label{eq:IOdynet}
\end{align}
where
\begin{align}
	T(z) \triangleq (I - G(z))^{-1}. \label{eq:Tdef}
\end{align}
Observe that the matrix $T(z)$ is nonsingular by construction. 

It is assumed that the input-output model $M(z)$ is known; the identification of $M(z)$ from  input-output (IO) data $\lbrace y(t), r(t) \rbrace$ is a standard  identification problem, provided the input signal $r(t)$ is sufficiently rich. The question of identifiability of the network is whether the network matrix $G(q)$ can be fully recovered from the  transfer matrix $M(z)$. We now give a formal definition of generic identifiability of the network matrix from the data $\lbrace y(t), r(t) \rbrace$ and from the graph structure.

\begin{definition}
(\cite{hendrickx-gevers-bazanella-identifiability-2019}) The network matrix $G(z)$ is generically identifiable from excitation signals applied to $\mathcal{B}$ and measurements made at $\mathcal{C}$ if, for any rational transfer matrix parametrization $G(P, z)$ consistent with the directed graph associated with $G(z)$, there holds
	\[
					C [I - G(P, z)]^{-1} B = C[I - \tilde{G}(z)]^{-1} B \implies G(P, z) = \tilde{G}(z),
	\]
	for all parameters $P$ except possibly those lying on a zero measure set in $\mathbb{R}^{N}$, where $\tilde{G}(z)$ is any network matrix consistent with the graph.
\end{definition}

%
In this paper, we discuss the identifiability  in terms of which nodes must be excited and/or measured in the subsets $\mathcal{B}$ and $\mathcal{C}$ in order to guarantee identifiability of the network. This approach is inspired by \cite{bazanella-gevers-hendrickx-etal-identifiability-2017} and a recent result in \cite{bazanella-gevers-hendrickx-network-2019}, which gives a necessary condition for generic identifiability of a network.
\begin{proposition}\label{corol1}
  The network matrix $G(z)$ is generically identifiable only if $\mathcal{B}, \mathcal{C} \neq \emptyset$, $\mathcal{F} \subset \mathcal{B}$, $\mathcal{S} \subset \mathcal{C}$  and $\mathcal{B} \cup \mathcal{C} = \mathcal{W}$.\label{cor:fromBandC}
\end{proposition}
As a consequence, every node of the network must be either excited or measured. Generic identifiability of a given network  can thus be equivalently characterized by the network's Excitation and Measurement Pattern, denoted EMP.
The concept of EMP, which led  to the concept of minimal EMP, was introduced in \cite{mapurunga-optimal-2021}. They are defined in the following. 
\begin{definition}
    A pair of selection matrices $B$ and $C$, with its corresponding pair of node sets $\mathcal{B}$ and $\mathcal{C}$,  is called an \emph{excitation and measurement pattern} - EMP for short.
  An EMP is said to be \emph{valid} if it is such that the network (\ref{eq:dynet1})-(\ref{eq:dynet2}) is generically identifiable.
  Let $\nu = |\mathcal{B}| + |\mathcal{C}|$ \footnote{$|\cdot|$ - Denotes the cardinality of a set.} be the cardinality of an EMP.
  A given EMP is said to be  \emph{minimal} if it is valid and there is no other valid EMP with smaller cardinality.
  \label{def:EMP}
\end{definition}
The following result establishes a lower and upper bound for the cardinality of a valid EMP for any network.
\begin{theorem}\label{cardinality}
The cardinality of a valid EMP for the identification of a dynamical network with $n$ nodes is at least equal to $n$ and at most equal to $2n-f-s$, where $f$ is the number of sources and $s$ the number of sinks.
\end{theorem}
\ni{\bf Proof.} The lower bound results from Proposition~\ref{corol1}; it can actually be achieved for trees and loops  \cite{bazanella-gevers-hendrickx-network-2019,mapurunga-gevers-bazanella-necessary-2022}.
As for the upper bound, we know by Proposition~\ref{corol1} that all sources must be excited and all sinks measured, while the remaining $n-f-s$ nodes must be excited or measured. Assuming that these are all excited and measured, then the cardinality of the EMP is $f+s+ 2(n-f-s)= 2n-f-s$.
 \cqfd
From now on, we drop the arguments $z$ and $t$ used in (\ref{eq:dynet1})-(\ref{eq:dynet2}) whenever there is no risk of confusion.

\section{Directed Acyclic Graphs and their properties}
\label{sec:DAG}

In this section we investigate the generic identifiability of dynamic networks whose topologies are associated with directed acyclic graphs, denoted DAG\footnote{For simplicity, we shall in the future just refer to a DAG rather than a network that has the topology of a DAG.}. These are very general classes of graphs, of which trees are a special case. We will derive necessary and sufficient conditions for the generic identifiability of these classes of networks and characterize which nodes need to be excited or/and measured in order to obtain a valid EMP for these networks.  Directed Acyclic Graphs are defined as follows.
\begin{definition}
 A directed acyclic graph is a directed graph that has no cycles.
    \label{def:DAG}
\end{definition}
A property of DAGs is that the sequence of their nodes can be relabeled   by a topological sorting algorithm \cite{kahn_topological_1962} in such a way that  $G_{ij}=0$ for $i<j$. In the sequel, we assume without loss of generality that the nodes of the dynamic networks we study in this section have been relabeled this way.
 As a result, the network matrix $G$ can be written as a lower triangular matrix.
 \begin{align}
         G =
          \left[
          \begin{matrix}
            0 & 0 & 0 & \ldots &  0 \\
            G_{21} & 0 & 0 &\ldots & 0 \\
            G_{31} & G_{32} & 0 & \ldots & 0 \\
						\vdots & \vdots & \ddots & \ddots & \vdots \\
            G_{n1} & G_{n2} & \ldots & G_{n,n-1} & 0
          \end{matrix}
          \right],
       	\label{eq:Glower}
				\end{align}
where some $G_{ij},   i > j$ are typically zero. 
In \cite{mapurunga-identifiability-2021} it was shown that for a network matrix (\ref{eq:Glower}) with all $G_{ij} \neq 0$ for $i > j$ generic identifiability is achieved if and only if all sources are excited, all sinks are measured, and every other node is 
both excited and measured.
Exciting and measuring all internal nodes is of course a very strong condition; we shall explain in the next section why it occurs when all $G_{ij}\neq 0$.

The following lemma establishes  relationships between such network matrix $G$ with the structure of a DAG and the corresponding I/O matrix $T$.

\begin{lemma}\label{lemma1}
Let $G$ be as in (\ref{eq:Glower}) and define $T=(I-G)^{-1}$. Then the following relationships hold.
\begin{align}
T_{ll} &= 1, \label{eq:Tdiag} \\
T_{lj} &= 0,\;\; \text{for}\; j > l, \label{eq:knownTdag} \\
T_{lj} &= \sum_{i = j}^{l-1} G_{li} T_{ij},\;\; \text{for}\; l >j \label{eq:recursiondag}\\
T_{lj} &= \sum_{i = j+1}^{l} T_{li} G_{ij},\;\; \text{for}\; l >j \label{eq:recursiondag2}\\
G_{lj} &= T_{lj} - \sum_{i = j+1}^{l-1} G_{li} T_{ij},\;\; \text{for}\; l >j  \label{eq:GTrel}\\
G_{lj} &= T_{lj} - \sum_{i = j+1}^{l-1} T_{li} G_{ij},\;\; \text{for}\; l >j  \label{eq:GTrel2}
\end{align}
\end{lemma}
\begin{proof}
The relations (\ref{eq:Tdiag})-(\ref{eq:recursiondag}) follow directly from $[ I - G] T= I_n$, while (\ref{eq:recursiondag2}) follows from $T [I-G] = I_n$. Observing that $T_{jj}=1$ in (\ref{eq:recursiondag}) yields (\ref{eq:GTrel}), while (\ref{eq:GTrel2}) follows similarly from (\ref{eq:recursiondag2}).
\end{proof}
In the sequel of this paper we shall illustrate all our results with the following 7-node DAG.\\

\begin{figure}[h!]
				\centering
				\includegraphics[width=0.5\textwidth]{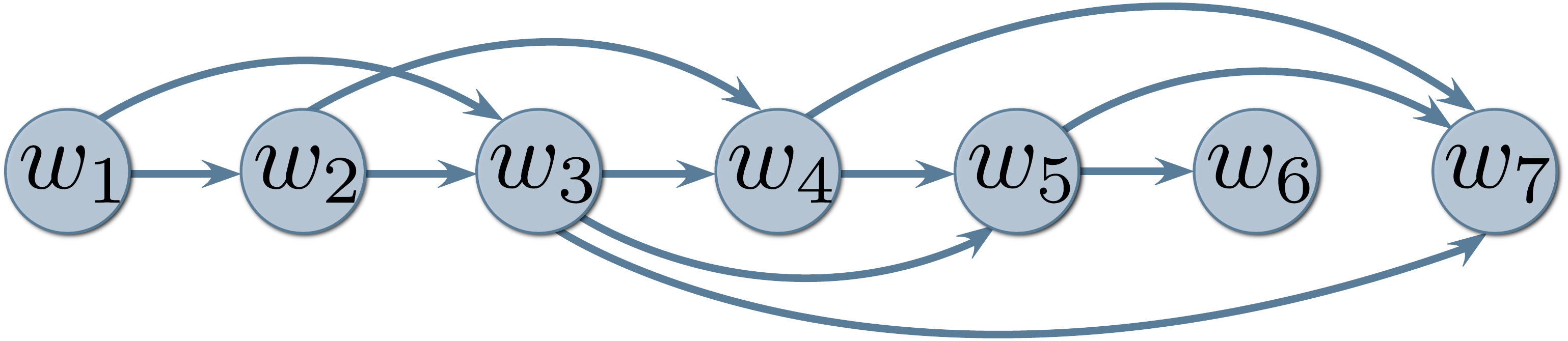}
				\caption{A 7-node DAG network}
				\label{fig:DAG7}
\end{figure}

For this network, equation (\ref{eq:GTrel2}) allows us, for example, to write:
\begin{align*}
G_{43}& = T_{43}\\
G_{53}&= T_{53} - T_{54}G_{43}  \\
G_{73}&= T_{73} - T_{74}G_{43} - T_{75} G_{53}
\end{align*}

Equations (\ref{eq:GTrel})-(\ref{eq:GTrel2}) define the expressions of the $G_{ij}$ recursively as a function of the $T_{ij}$ and of the previously computed $G_{kl}$. But the $G_{ij}$ can also be expressed explicitly as functions of the $T_{ij}$ as is shown in the following Theorem, which is one of the main results of this paper. 
It shows that for DAGs one can compute the unknown transfer functions $G_{ij}$ explicitly as a function of the elements of the I/O matrix.

\begin{theorem}\label{lem:GijSij}
Define $S \bydef T^{-1}$ with elements $S_{ij}$. Then the following results hold:\\
(1) $G_{lj}= -S_{lj}$, and hence $S_{lj}=0$ for each pair $\lbrace l,j \rbrace$, $l \neq j$, 
for which $G_{lj}$ is known to be zero;\\
(2) $G_{lj}$ can be written as a sum of products of $T_{ki}$ with $k\leq l$ and $i\geq j$. 
\end{theorem}
\ni{\bf Proof.} Item (1) follows from $G= I - T^{-1}$ and the fact that the inverse of a lower triangular matrix with `ones' on its diagonal is a lower triangular matrix with `ones' on its diagonal. \\
Item (2) (in particular the fact that $T^{-1}$ does not contain any $T^{-1}_{ij}$) is a property of the inverse of a lower  triangular matrix with `ones' on its diagonal. It  follows from the direct computation of $T^{-1}$, but it also follows by substituting the $G_{ij}$  on the right hand side of (\ref{eq:GTrel2}) by their expressions computed from the same equation. 
\cqfd
To illustrate the result (2) of Theorem~\ref{lem:GijSij}, we observe that if we substitute $G_{53}$ and $G_{43}$ in the expression of $G_{73}$ above by their expressions, we obtain
\begin{align*}
G_{73}&= T_{73} - T_{74}T_{43} - T_{75} T_{53} + T_{75} T_{54} T_{43}
\end{align*}
as claimed.

Not only does Theorem~\ref{lem:GijSij} provide an explicit expression for the $G_{lj}$ as a function of the input-output elements $T_{lk}$ but it also proves very useful to establish relations among the $T_{lk}$. Indeed, for each $G_{lj}$ that is known to be zero, result (2) of the Theorem allows one to compute one of the $T_{lk}$ as a function of others. This may then allow one to eliminate the need for the excitation or measurement of some nodes as we shall show in Section~\ref{EMP1DAG}. Returning again to the 7-node DAG of Figure~\ref{fig:DAG7}, since $G_{52}=0$, it follows that $S_{52}=0$, which implies
\begin{align*}
T_{52}&=    T_{53} T_{32} + T_{54}T_{42} + T_{54} T_{43} T_{42}.
\end{align*}

\section{Necessary conditions for the identification of a DAG}\label{neccondDAG}

It was shown in \cite{bazanella-gevers-hendrickx-network-2019} that a necessary condition for the generic identifiability  of any network is that all sources must be excited, all sinks must be measured, and that each other node must be either excited or measured. Here we show that, for DAGs, some additional necessary conditions are required for two special classes of nodes, namely the dources and the dinks that were defined in Section~\ref{sec:Problem}. Our main result  is the following.

\begin{theorem}
Consider a dynamic network with the  topology of a directed acyclic graph whose network matrix is given in (\ref{eq:Glower})  with some known $G_{ij} = 0$ for $i > j$. Then this network is generically identifiable only if the following conditions hold.\\
(1) each node is either excited or measured;\\
(2) all sources are excited and all sinks are measured;\\
(3) all  dources are excited and all dinks are measured.
\label{theo:dag2}
\end{theorem}	
\ni{\bf Proof.} 
Conditions (1) and (2) have been shown to be necessary for the idenfication of any network in \cite{bazanella-gevers-hendrickx-network-2019}.\\
We  turn to item (3). We first prove that the excitation of all dources is necessary. Consider a node $l$ that is an outneighbor of $i$ such that all in-neighbors of node $i$ are connected to that outneighbor $l$. This means that node $i$ is a dource. It then follows that for each in-neighbor $j$ of node $i$, we have
\begin{itemize}
\item $G_{ij} \neq 0$ since $j$ is an in-neighbor of $i$;
\item  $G_{lj}\neq 0$ by the assumption above.
\end{itemize}
To show that the dource $i$ needs to be excited, we focus on the transfer function $T_{li}$ between the dource $i$ and the out-neighbor $l$ that is connected to all in-neighbors of $i$. It follows from (\ref{eq:GTrel2}) that in the equations defining the  $G_{ij}$ as a function of the $T_{lk}$ and the other $G_{ij}$, this specific transfer function $T_{li}$ appears only in the expressions of $G_{lj}$ in which  $l$ is the considered out-neighbor of dource $i$, and $j$ is either the dource $i$ itself or one of its $\mathcal{N}{_i^-}$ in-neighbors. We now observe that in these  $1+\mathcal{N}{_i^-} $ equations,  $T_{li}$ is either  multiplied by '1' or by the $G_{ij}$ that relate the in-neighbors of dource $i$ to itself. These $G_{ij}$ are nonzero (see above).  Each  of these equations $1+\mathcal{N}{_i^-}$ equations therefore contains an unknown $G_{lj}$ on the left hand side and the unknown $T_{li}$ on the right hand side. 
Hence node $i$ must be excited in order to compute $T_{li}$ and therefore the  edges $G_{lj}$ linking the dource $i$ and all its in-neighbors  to this particular out-neighbor.\\
The proof for the necessity of measuring all dinks is the exact dual of the proof for the necessity of excting all dources and will therefore be omitted.
\cqfd
We illustrate the result of Theorem~\ref{theo:dag2} with the 7-node DAG network pictured in Figure~\ref{fig:DAG7}. This network has one source (node 1), two dources (nodes 2 and 5), two sinks (nodes 6 and 7) and one dink (node 4). 

We first show why node $5$ must be excited. It is a dource because all its in-neighbors (nodes $3$ and $4$)  are connected to out-neighbor $7$. We write the equations (\ref{eq:GTrel2}) that relate the out-neighbor to the dource and to its two in-neighbors.
\begin{align*}
G_{75}& = T_{75}\\
G_{74}&= T_{74} - T_{75} G_{54}  \\
G_{73}&= T_{73} - T_{74}G_{43} - T_{75} G_{53}
\end{align*}
We observe that, even if all quantities other than $T_{75}$ on the right hand side were known (i.e. $T_{74}, T_{73}, G_{54}, G_{43}, G_{53}$), it would be impossible to identify the transfer functions $G_{75}, ~G_{74}$ and $G_{73}$ without knowing $T_{75}$, which requires exciting node 5.
We add that, if an edge $G_{76}$ was added, the first equation would be replaced by $G_{75} = T_{75}- G_{76}T_{65}$, and the conclusion would thus be identical.


We now illustrate why node 4 must be measured. It is a dink because it has an in-neighbor (node 3) that is connected to all its out-neighbors (nodes 5 and 7). We write the equations (\ref{eq:GTrel}) that relate its in-neighbor 3 to the dink and to all  its out-neighbors.
\begin{align*}
G_{43}& = T_{43}\\
G_{53}&= T_{53} - G_{54}T_{43}  \\
G_{73}&= T_{73} - G_{74}T_{43} - G_{75} T_{53}
\end{align*}
We observe that $T_{43}$ is required to compute $G_{43}$, and that it cannot be computed from the other two equations for the same reasons as above. Thus, node 4 must be measured for the identification of $G_{43}$, $G_{53}$ and $G_{73}$.

Notice that a given node can be both a dource and a dink. In a ``full'' DAG as in (\ref{eq:Glower}), with all $G_{ij} \neq 0$,  all internal nodes are dources and dinks, which explains the need for exciting and measuring them all, as shown in  \cite{mapurunga-identifiability-2021}.

\section{Constructing a valid EMP for a DAG}\label{EMP1DAG}

In this section, we show how to construct a valid EMP while trying to keep the cardinality of this EMP low.
Recall that this cardinality is always between $n$ and  $2n-f-s$. 

An easy solution for the construction of a valid EMP results from Theorem~\ref{lem:GijSij}. Indeed, each $G_{ij}$ can be expressed explicitly as a combination of $T_{kl}$. The collection of all these $T_{kl}$ indicates which node excitations and which node measurements will lead to a valid EMP. In order to reduce the cardinality of this explicit solution, one can, in a second step, take advantage of the possible replacement of some of these $T_{kl}$ by others using the equations $S_{ij}=0$ for each $G_{ij}$ that is known to be zero: see item (1) of Theorem~\ref{lem:GijSij}.


For the 7-node DAG of Figure~\ref{fig:DAG7} the explicit solution provided by Theorem~\ref{lem:GijSij} yields the following solution for the $G_{ij}$: 
\begin{align*}
G_{21}&=T_{21}\\
G_{31}&= T_{31}-T_{21}T_{32}, G_{32}=T_{32}\\
G_{42}&= T_{42} - T_{32}T_{43}, G_{43}= T_{43}\\
G_{53}&= T_{53}-T_{43}T_{54}, G_{54}=T_{54}\\
G_{65}&= T_{65}\\
G_{73}&= T_{73}- T_{75}T_{53} - T_{74}T_{43} + T_{75}T_{54}T_{43}\\
G_{74}&=T_{74} - T_{54}T_{75}, G_{75}= T_{75}
\end{align*}
Collecting all the indices that appear as inputs and outputs of the $T_{kl}$ yields the valid EMP ${\cal B} = \{1, 2, 3, 4, 5 \}$ and ${\cal C} = \{2, 3, 4, 5, 6, 7 \}$. It has cardinality 10. Observe that nodes 2, 3, 4, 5 are both excited and measured. According to Theorem~\ref{theo:dag2}, nodes 2 and 5 are dources and node 4 is a dink. Hence, nodes 2 and 5 must be excited, as well as node 1, which is a source; while node 4 must be measured, as well as nodes 6 and 7 which are sinks. Starting from this initial EMP with cardinality 10, one can then use the equations of Theorem~\ref{lem:GijSij} and Lemma~\ref{lemma1} (in particular the equations $S_{ij}=0$) to check whether one can eliminate any one of nodes 3 and 4 from ${\cal B}$, or any one of nodes 2, 3, 5 from ${\cal C}$.
However, such  procedure is tedious, and there appears to be no systematic way to proceed with the elimination of excitations or measurements. Therefore, we propose a recursive procedure for the construction of a valid EMP which is based on Theorem~\ref{lem:GijSij} and Lemma~\ref{lemma1}. \\

\ni {\bf Recursive procedure for the construction of a valid EMP.}\\
First build the matrix $S$, replacing each $S_{lj}$ by $-G_{lj}$, where these $G_{lj}$ are computed using equation (\ref{eq:GTrel2}) of Lemma~\ref{lemma1}. For each $G_{lj}$ that is known to be zero, equate the zero element in $S_{ji}$ to the expression resulting from the same equation (\ref{eq:GTrel2}).

Now construct the preliminary EMP with the excitations and measurements required by the structure of the DAG: sources and dources must be excited, dinks and sinks must be measured. Call it \emp{0}, thus defining a ${\cal B}_0$ and a  ${\cal C}_0$. List the corresponding known $T_{ij}$, i.e. all $T_{ij}$ for which $j \in {\cal B}_0$ and $i \in {\cal C}_0$. 

\begin{table*}[!t]
			  \centering
				\caption{Construction of an EMP column-wise.}
			
				\label{tab:column}
				\begin{adjustbox}{width=0.82\linewidth}
				\begin{tabular}{| c | c| c | c | c |}
				\hline
																column & Added E or M & EMP update  & Known $G_{ji}$ & Known $T_{ji}$ \\
								\hline
								0 & -- &E125, M467 & -- & $T_{41}, T_{42}, T_{61}, T_{62}, T_{65}, T_{71}, T_{72}, T_{75}$\\
								1& M23 & E125, M23467 & $G_{21}, G_{31}, G_{43}$ &  + $T_{32}, T_{43}, T_{63}, T_{73}$ \\
								2& -- & E125, M23467 & $G_{32}, G_{42} $ & + $T_{64}$, $T_{74}$ \\
								3& -- & E125, M23467 & $G_{53}$ from $S_{63}$, $G_{73}$ & --\\
								4 & -- & E125, M23467 & $G_{54}$ from $S_{64}$, $G_{74}$ & -- \\
								5& -- & E125, M23467 & $G_{65}, G_{75}$ & -- \\
								\hline
				\end{tabular}
				\end{adjustbox}
\end{table*}

Now  proceed stepwise within the columns of $S$, say from column 1 to column $n$, as explained below.

\indent 1)~Column 1: the unknown $G_{j1}$ appear in the first column of $S$. Add to \emp{0} whatever excited node  or measured node  is required to be able to identify all $G_{j1}$. There may be several choices. Use the remaining $S_{j1}=0$  equations of column 1 to compute new elements $T_{kl}$. Update \emp{0} to \emp{1}, update the known $G_{j1}$, and update the known $T_{kl}$.
\\
\indent 2)~Column 2: the unknown $G_{j2}$ appear in the second column of $S$. Add to \emp{1} whatever excited node  or measured node  is required to be able to identify all $G_{j2}$. There may be several choices. Use the remaining $S_{j2}=0$   equations of column 2 to compute new $T_{kl}$. Update \emp{1} to \emp{2}, update the known $G_{j2}$, and update the known $T_{kl}$.
\\
\indent 3)~Continue until all columns of $S$ containing elements $G_{lj}$ have been covered.

%
%

%

Before we illustrate this procedure with our Example, let us make the following comments.

\ni{\bf Comments}
\begin{itemize}
\item The computation of elements $G_{lj}$ based on column $j$ of $S$ may require that several elements of that column be used jointly, leading to the solution of a linear system of equations. One must check that these equations are linearly independent. 
\item The procedure proposed above uses a column by column approach, going from left to right. Other approaches can be used, such as covering the columns from right to left, using a row by row approach, etc.  These different approaches will typically lead to different valid EMPs.
\end{itemize}
\vspace{3mm}
\ni {\bf The EMP  procedure applied to the DAG of Figure  1.}\\
For brevity of notation, for this example with 7 nodes,  we shall represent an EMP that has ${\cal B} = \{1, 2, 4, 5 \}$ and ${\cal C} = \{2, 3, 4, 6, 7 \}$, say, by E1245, M23467. 

For this example, we know a priori that nodes 1, 2 and 5 must be excited, being a source and two dources, while nodes 4, 6 and 7 must be measured, being a dink and two sinks. As a result, the starting EMP, denoted \emp{0} above, is E125, M467. 
With this initial \emp{0}, the elements $T_{41}, T_{42}, T_{61}, T_{62}, T_{65}, T_{71}, T_{72}, T_{75}$ are known. 

In Table I describing the procedure, we have listed the a priori information as being in column $0$. 
We observe that the  elements of column 1 can be identified by  the addition of M23, i.e. the measurement of nodes 2 and 3.  It turns out that with this addition of M23, all other elements $G_{lj}$ can subsequently be computed, and the procedure ends with a valid EMP defined as E125, M23467, which has  cardinality  8.

An alternative is to add E3 for the identification of the elements of column 1, and E4 for the identification of the elements of column 2. This yields the alternative valid EMP defined as E12345, M467, with  the same cardinality 8.
A row by row procedure, from top to bottom, applied to the same example leads to the  valid EMP: E125, M234567, which has cardinality 9. In comparison, the explicit solution based on the decomposition of each $G_{lj}$ as a function of the $T_{kl}$ only, as explained above, has cardinality 10.

\section{Conclusions}\label{Conclusion}
We have pursued our study of specific substructures of dynamic networks, with the aim of designing EMPs that guarantee identifiability for these specific substructures. 
In \cite{hendrickx-gevers-bazanella-identifiability-2019} a necessary and sufficient condition was established for trees, with a corresponding valid (and minimal) EMP. 
In \cite{mapurunga-identifiability-2021} necessary and sufficient conditions for network structures with parallel paths were derived. In \cite{mapurunga-gevers-bazanella-necessary-2022} it was shown that any loop with more than 3 nodes can also be identified with a minimal EMP and the design of such loops was presented. In the present paper we have focused on networks that have the structure of a DAG. We have shown that DAGs have specific properties which facilitate the design of valid EMPs. A rather surprising result of our analysis has been to show that two specific and well-defined nodes, which we have called dources and dinks, have specific excitation and measurement requirements. Dources, just like sources, must be excited and dinks, just like sinks, must be measured in order for a DAG network to be identifiable. With these constraints under  our belt, we have produced a recursive EMP design procedure that takes advantage of the properties of DAGs.

\bibliographystyle{IEEEtran}
\bibliography{StrucDyNetB}

\end{document}